# Development of a Machine-Learning System to Classify Lung CT Scan Images into Normal/COVID-19 Class

**Seifedine Kadry[1], Venkatesan Rajinikanth [2], Seungmin Rho[3,\*], Nadaradjane Sri Madhava Raja[2], Vaddi Seshagiri Rao[4], Krishnan Palani Thanaraj [2]**

[1]Department of Mathematics and Computer Science, Faculty of Science, Beirut Arab University, Lebanon,
[2]Department of Electronics and Instrumentation Engineering, St. Joseph's College of Engineering, Chennai 600119, India;
[3]Department of Software, Sejong University, Seoul 05006, Korea;
[4]Department of Mechanical Engineering, St. Joseph's College of Engineering, Chennai 600 119, India;

**Abstract:** Recently, the lung infection due to Coronavirus Disease (COVID-19) affected a large human group worldwide and the assessment of the infection rate in the lung is essential for treatment planning. This research aims to propose a Machine-Learning-System (MLS) to detect the COVID-19 infection using the CT scan Slices (CTS). This MLS implements a sequence of methods, such as multi-thresholding, image separation using threshold filter, feature-extraction, feature-selection, feature-fusion and classification. The initial part implements the Chaotic-Bat-Algorithm and Kapur's Entropy (CBA+KE) thresholding to enhance the CTS. The threshold filter separates the image into two segments based on a chosen threshold 'Th'. The texture features of these images are extracted, refined and selected using the chosen procedures. Finally, a two-class classifier system is implemented to categorize the chosen CTS (n=500 with a pixel dimension of 512x512x1) into normal/COVID-19 group. In this work, the classifiers, such as Naive Bayes (NB), k-Nearest Neighbors (KNN), Decision Tree (DT), Random Forest (RF) and Support Vector Machine with linear kernel (SVM) are implemented and the classification task is performed using various feature vectors. The experimental outcome of the SVM with Fused-Feature-Vector (FFV) helped to attain a detection accuracy of 89.80%.

## 1. Introduction

The abnormality/infection in the internal body organs are more acute compared to other diseases. Further, the diseases in internal organs are commonly prescreened using the non-invasive methods such as bio-signals and bio-images [1-3]. The infection in lung due to the climatic condition and microorganisms are very common in humans and this infection may cause various symptoms ranging from caught, cold, fever and mild/severe pneumonia [4-6].

The respiratory tract infection due to the Coronavirus Disease (COVID-19) is emerged as one of the major threat globally due to its acuteness and the infection rate. It is one of the major communicable infectious diseases caused by Severe Acute Respiratory Syndrome-Corona Virus-2 (SARS-CoV-2) and according to a recent report [7,8], it affected a larger human community, irrespective of their race and gender. The infection caused by COVID-19 severely affects the respiratory system by causing the severe pneumonia. Due to its harshness and the spreading rate, the World Health Organization (WHO) recently announced it as pandemic [9]. Even though various controlling and treatment procedures are implemented from December 2019 to till date, the mortality due to COVID-19 infection is rapidly increasing.

Due to its acuteness, a considerable number of already initiated to discover the possible solution for the problem due to COVID-19. The earlier research works are related to; (i) The succession and prediction of the COVID-19 to alert the people [10-12], (ii) Precautionary measures to be implemented to control the spread [13,14], (iii) Exploring the structure of the virus to find the solution and Clinical level handling of the pneumonia caused by COVID-19 [15-22].

Among the above said procedures, the clinical level handling gets the priority, in which a possible treatment practices are suggested and implemented by the doctors to control and cure the infection in respiratory tract using various procedures. The clinical level detection of COVID-19 requires two accepted methodologies [23,24]; (i) Clinical level testing using Reverse Transcription-Polymerase Chain Reaction (RT-PCR) test to confirm the disease, and (ii) Image assisted procedure to identify the severity of pneumonia in lungs using Computed-Tomography scan Slices (CTS) and/or chest radiograph (Chest X-ray) assisted diagnosis. During the RT-PCR test, if the result is negative, then the person is a normal and when the RT-PCR provides a positive result, then the person is immediately admitted in hospital for further treatment. When the patient is admitted, the doctor will suggest a combination of treatment procedures, including the image assisted treatment, which helps to choose the possible drug and its dosage level [25]. During this practice, all the RT-PCR positive patients are initially screened with CTS/Chext X-ray to identify the severity of the lung infection. When the treatment is initiated, the patient will be treated till the respiratory system functions well [13].

The treatment planning and the implementation will become a challenging task, when large number of COVID-19 infected patients is admitted in the hospital. Furthermore, the diagnostic burden also rises due to the mass screening of the patients with the imaging modality. Hence, to reduce the burden of the doctors, a computer assisted methods are essential for the initial diagnosis and based on the outcome by the computerized procedure, the doctors can plan and execute the treatment. Normally, the computer based detection procedure can be executed by a skilled technician or a doctor and the findings of this procedure can be shared to the pulmonologist for further assessment. Recently, a considerable number of CTS assisted detection procedures for COVID-19 is reported in the literature [26-33]. Every procedure considered the axial/coronal view of the CTS and the most of the procedures are interested in extracting the pneumonia lesion from the infection to assess the severity of the disease. Still there is a need for a considerable number of image examination procedures, which can be used in future for the clinical level practice.

This research proposes a Machine-Learning-System (MLS) to classify the CTS into normal/COVID-19 category using a sequence of methods. The procedures implemented in this MLS is as follows; (i) Tri-level thresholding with Chaotic-Bat-Algorithm and Kapur's Entropy (CBA+KE), (ii) Separation of the image into Region-Of-Interest (ROI) and artifact by a threshold filter, (iii) Feature extraction using chosen methodology, (iv) Feature ranking and selection based on statistical test, (v) Implementation of serial fusion to get the one-dimensional Fused-Feature-Vector (FFV) and (vi) Classifier implementation and validation.

The proposed work is experimentally investigated using MATLAB® software and the essential CTS are collected from the available benchmark datasets. In this work, 500 images (250 normal and 250 COVID-19) of dimension 512x512x1 pixels are utilized for the evaluation. This work implemented a five-fold cross validation during the classification task and the best value attained is considered as the finest result. Further, the proposed work also presents a performance evaluation of the classifiers, such as Naive Bayes (NB), k-Nearest Neighbors (KNN), Decision Tree (DT), Random Forest (RF) and Support Vector Machine (SVM) using the chosen feature vector. The propose MLS helped to achieve a classification accuracy of 81.20% (SVM), 85.80% (RF), and 89.80% (SVM) for various features employed to train, test and validate the classifiers.

This study is prearranged as follows; section 2 presents context and section 3 shows the methodology. Section 4 summarizes the experimental outcome and its discussions. Section 5 describes the conclusion.

**2. Context**

COVID-19 is a recently emerged infectious disease discovered initially in China (Wuhan) in December 2019 [9,14]. The drug discovery for this disease is still in the research phase and no approved drug is available for COVID-19. Due to these reasons, the mortality rate is rising globally [7].

In recent days, a number of image assisted detection procedure for COVID-19 is discussed by the researchers and Table 1 present the summary of few recent techniques.

**Table 1.** The summary of image based COVID-19 detection procedure

| Reference | Implemented investigative procedure | Image modality | Findings |
|---|---|---|---|
| Rajinikanth et al. [26] | Harmony-Search and Otsu's based image thresholding and watershed-Segmentation is implemented to extract COVID-19 infection. | CT | Disease severity prediction based on the size of the infection with respect to the lung is discussed |
| Rajinikanth et al. [27] | Firefly and Shannon's entropy based image thresholding and Markov-random-field segmentation is implemented to extract COVID-19 infection. | CT | This work provided a segmentation accuracy of >92% during the COVID-19 lesion extraction |
| Wu et al. [28] | This work implemented a deep-learning procedure for the segmentation and classification of COVID-19 infection from CT images attained from 200 patients. | CT | Provided the dice score of 78.3% for segmentation. Further, helped to achieve an average sensitivity of 95.0% and a specificity of 93.0% during the classification. |
| Khan et al. [29] | A deep-learning based diagnosis of COVID-19 is implemented using CoroNet architecture. | Chest X-Ray | This work provided a classification accuracy of 89.50%. |
| Rahimzadeh and Attar [30] | Implements a modified deep convolutional neural network for the COVID-19 diagnosis. | Chest X-Ray | This work provided a classification accuracy of 99.56% for the disease class and average accuracy of 91.4%. |
| Ozkaya et al. [31] | This work implemented a deep-learning based on features fusion and ranking technique. | CT | This work helped to attain better values of accuracy (98.27%), sensitivity (98.93%), specificity (97.60%), precision (97.63%), and F1-score (98.28%). |
| Zhou et al. [32] | U-Net with attention mechanism is implemented to segment the COVID-19 infection. | CT | This work provided the following measures; Dice Score=69.1%,Sensitivity=81.1% and Specificity = 97.2%, |

The details discussed in Table 1 presents recently implemented methodologies to segment and detect the COVID-19 infection using the CT and chest X-ray images. Further, a detailed review of the image assisted procedures existing in the literature can be found in the recent work of Shi et al. [33]. From these earlier works, it can be noted that, the image assisted COVID-19 detection system is essential to support the doctor during the disease diagnosis task. Hence, in this research work, a MLS is proposed to detect the disease using the CTS.

## 3. Methodology

This part of the work presents the methodology implemented in this system which can work well on the CTS of the views, such as axial, coronal and sagittal. For experimental demonstration, only the axial-view of the CTS is considered. The various stages employed in the proposed work are clearly depicted in Figure 1.

The infected patient primarily assed with a radiology assisted imaging procedure (CT scan), which provides a reconstructed three-dimensional (3D) image of the respiratory tract. Assessment of the 3D requires complex computations and hence, the 3D images are separated into 2D slices during the examination. In this work, the axial CTS of normal/COVID-19 class are considered to test the performance of the proposed MLS. Initially, the visibility of the infected section is enhanced using a tri-level thresholding implemented using the Chaotic-Bat-Algorithm and Kapur's entropy (CBA+KE). After the thresholding, a bi-level threshold filter discussed in [26,34] is implemented to separate the image into Region-Of-Interest (ROI) and artifact. A feature extraction procedure is then implemented to extract the image features from original, threshold and ROI. After extracting the features; the dominant features from each image category is selected using the statistical test and the chosen features are then considered to train, test and validate the classifier system implemented in this work. Further, a future fusion technique is also employed to increase the classification accuracy.

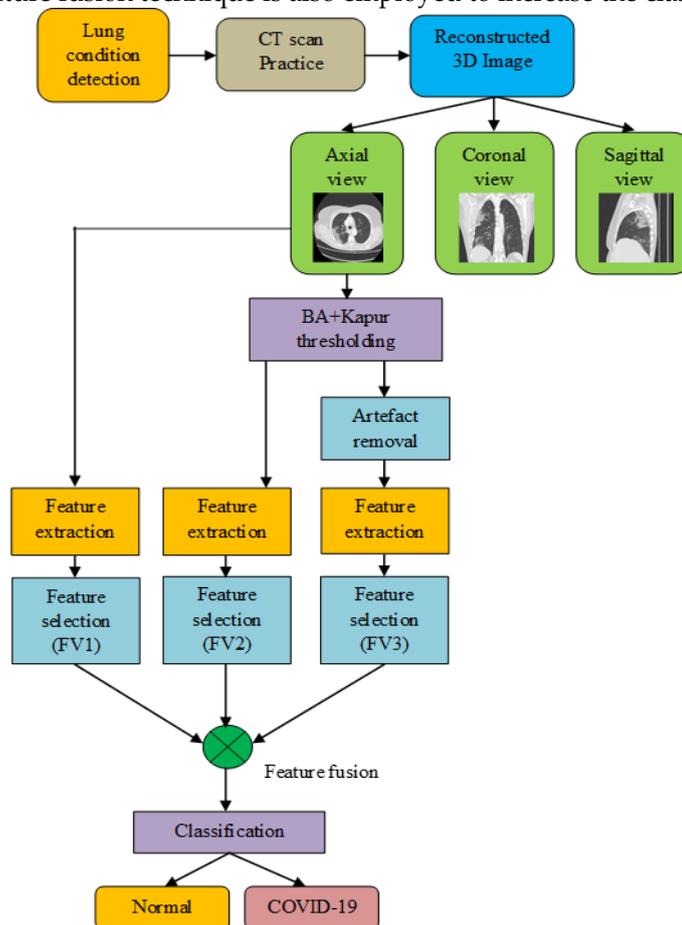

**Figure 1.** Proposed Machine-Learning system to recognize the COVID-19 from CT scan images

### 3.1 Image multi-thresholding

Image thresholding is one of the widely adopted enhancement technique considered to improve the visibility of grayscale/RGB images. In this work, the Kapur's entropy thresholding discussed in [35-37] is implemented to enhance the CTS for further assessment.

The mathematical description of the Kapur's entropy is discussed blow;

Let us consider a chosen dimension of the grayscale image with $L$ gray-levels ($0$ to $L-1$) with a total pixel value of $G$. If $F(i)$ denotes the frequency of the i$^{th}$ intensity-level; then the pixel distribution of the image will be;

$$G = F(0) + F(1) + ... + F(L-1) \quad (1)$$

Then the probability of i$^{th}$ intensity-level is represented by;

$$P(i) = F(i)/G \quad (2)$$

If there are $T$ thresholds as: $(t_1, t_2, ..., t_T)$, where $1 \leq T \leq L-1$.

During the thresholding operation, the image pixels are separated into $T+1$ groups based on the assigned threshold value. After separating the images as per the chosen threshold, the entropy of each group is computed separately and combined to get the final entropy.

For a tri-level threshold problem, the computed entropy will be;

$$F(t_1, t_2, t_3) = E_0 + E_1 + E_2 \quad (3)$$

$$E_0 = -\sum_{i=0}^{i=t_1-1} \frac{P_i}{\omega_0} \ln \frac{P_i}{\omega_0}, \omega_o = \sum_{i=0}^{i=t_1-1} P_i$$

$$E_1 = -\sum_{i=t_1-1}^{i=t_2-1} \frac{P_i}{\omega_1} \ln \frac{P_i}{\omega_1}, \omega_1 = \sum_{i=t_1-1}^{i=t_2-1} P_i \quad (4)$$

$$E_2 = -\sum_{i=t_2-1}^{i=t_3-1} \frac{P_i}{\omega_2} \ln \frac{P_i}{\omega_2}, \omega_2 = \sum_{i=t_2-1}^{i=t_3-1} P_i$$

where $E$ =entropy, $P$ =probability distribution, and $\omega$ =probability occurrence.

During this operation, the objective is to find; $Kapur_{max} = F_{Kapur}(T) = F(t_1, t_2, t_3)$ (5)

In this research, identification of $F_{Kapur}(T)$ is achieved by the CBA. In the literature, a number of procedures are implemented to enhance the optimization performance of the bat-algorithm (BA) and in the proposed work, the search operator in the traditional BA is improved using the Lorenz-Attractor ($\psi$) discussed in [38,39].

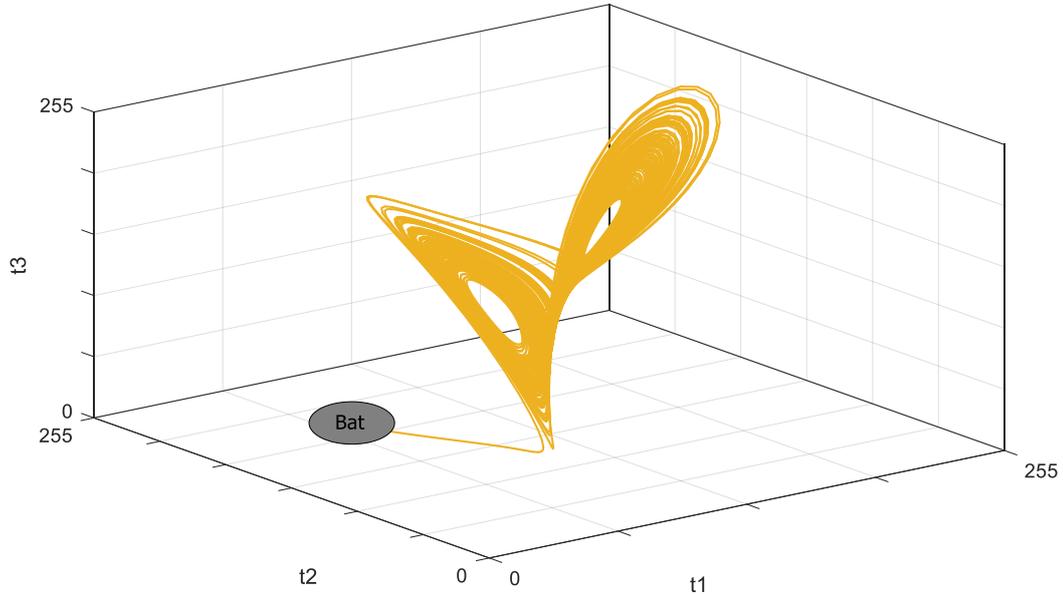

**Figure 2.** Search pattern made by a single bat to find $t_1, t_2, t_3$ using the Lorenz-attractor

The BA bas the following representations [40-42]:

Velocity update = $v_j^{n+1} = v_j^t + [p_j^t - G_{best}]F_j$ (6)

Location update = $p_j^{n+1} = P_j^n + v_n^{n+1}$ (7)

Frequency alteration = $f_j = f_{min} + (f_{max} - f_{min})\lambda$ (8)

where $\lambda$ is a random value of range [0,1].

Eqn. (8) drives Eqn. (6) and Eqn. (7) and hence, the choice of the frequency value should be appropriate. Updated value for every bat is produced based on; $p_{new} = p_{old} + (\sigma^n * \psi)$ (9) where $\psi$ is a Lorenz-Attractor and $\sigma$ =loudness constraint.

The expression of the loudness variation can be represented as;

$$\sigma_j^{n+1} = \alpha \sigma_j(n) \quad (10)$$

where $\alpha$ is a variable with a value $0<\alpha<1$.

The typical search of a bat in the three-dimension search space is depicted in Figure 2. Every bat is responsible to find $F(t_1, t_2, t_3)$ for the considered grayscale image and the pseudo-code clearly describes the proposed thresholding process.

The pseudo-code for CBA+KE thresholding;

---

*Step1*: Initialise the CBA with following parameters {number of bats=25, search dimension=3, objective function= $F_{Kapur}(T)$, total iteration=3000 and stopping criteria=maximal iteration, $f_{min} = 0$ and $f_{max} = 50$ and varies in steps of 0.05.

*Step2*: Randomly initialize the bats in the 3D search space and compute $F_{Kapur}(T)$ for each bat.

*Step3*: Find the $G_{best}$ attained by a bat and update the velocity and position using; $v_j^{n+1} = v_j^t + [p_j^t - G_{best}]F_j$

$$p_j^{n+1} = P_j^n + v_n^{n+1}$$

*Step4*: When the search iteration rises, update the position of every bat using;

$$p_{new} = p_{old} + (\sigma^n * \psi)$$

*Step5*: Is maximal iteration is reached (or) all the agents attained $F_{Kapur}(T)$?

If yes, stop the search and declare the thresholds $t_1, t_2, t_3$.

Else, repeat steps 2 to 4, till maximal iteration is attained.

---

### 3.2 Image separation

The accuracy of the disease detection using bio-images depends mainly on the quality of the image considered. The lung CTS is normally associated with the lung section to be examined along with other unwanted section, such as the bone segment and other body parts. In order to have a better diagnosis using the computer assisted procedures, it is necessary to consider the Region-Of-Interest (ROI) from the medical image. In this work, a threshold filter implemented in [34] is considered to separate the threshold image into ROI and artifact. As discussed by Rajinikanth et al. [26], the threshold level (*Th*) of the filter is initially identified manually and this threshold is then considered for all other images. The extracted ROI has the pneumonia infection section due the COVID-19 and this section is then considered for further assessment. From the ROI, the pneumonia infection is then segmented using the watershed-segmentation discussed in [26]. The segmentation result confirms that, proposed methodology helped to extract the pneumonia infected region from the axial, coronal and sagittal view of the CTS.

### 3.3 Feature extraction and selection

All the images (original, threshold, and ROI) considered in this work are in 2D form and hence, the 2D image feature extraction procedures, such as Discrete Wavelet Transform (DWT), Gray-Level Co-Occurrence Matrix (GLCM) and Hu Moments (HuM) are implemented. Further, the entropy features, such as Kapur, max [43-45], Renyi [46,47], Tsallis [48], Shannon [49], Vajda, and Yager [50,51] are also extracted and considered as the prime features.

- **DWT:** It evaluates the non-stationary details in image and the arithmetical expression of DWT is indicated as follows;

When a wavelet has the function $\psi(t) \in W^2(r)$, then its DWT will be;

$$DWT(a,b) = \frac{1}{\sqrt{2^a}} \int_{-\alpha}^{\alpha} x(t)\psi^*\left(\frac{t-b2^a}{2^a}\right)dt \qquad (11)$$

where $\psi(t)$ is the principle wavelet, the symbol '*' denote the complex conjugate, a and b ($a,b \in R$) are scaling parameters for image dilation and transition correspondingly.

The proposed work extracts 40 numbers of the features using DWT [43]. After extracting these features from the normal/Covid-19 class images, student's t-test based statistical evaluation is executed and these features are ranked based on the attained t-value and the DWT features whose p-value is >0.05 is discarded. This feature selection procedure helped to attain 13 numbers of one-dimensional feature vector and these are considered and the dominant DWT features.

Let the feature vector (13x1) attained from this procedure be;

$$f_{DWT} = (f'_{a1}, f'_{a2}, ..., f'_{a13}) \qquad (12)$$

- **GLCM and HuM:** In the image processing literature, a considerable number of research works are considered the GLCM features during the image recognition and the categorization tasks. The implementation of the HuM for the lung CTS and Chest X-ray is already discussed in the recent work of Bhandary et al. [34]. In this work, 18 numbers of the GLCM and 9 numbers of the HuM are considered as the dominant features.

$$f_{GLCM} + f_{HuM} = (f'_{b1}, f'_{b2}, ..., f'_{b18}) + (f'_{c1}, f'_{c2}, ..., f'_{c9}) \qquad (13)$$

- **Entropy features:** Entropy is the measure of the abnormality existing in the image and this feature provides the essential information on the lung abnormality in the CTS. In this work, 7 entropy features are considered and the essential information of these features can be found in [43-47].

$$f_{Entropy} = (f'_{d1}, f'_{d2}, ..., f'_{d7}) \qquad (14)$$

For a given image, the 1D Feature-Vector (FV) can be obtained by combining and sorting the dominant features, such as $f_{DWT}$, $f_{GLCM}, f_{HuM}$, and $f_{Entropy}$. Feature extraction is separately implemented on the three images cased and the attained Feature-Vectors (FV) with a size of 47 features is arranged as follows;

Original image= $FV_1 = f_{DWT} + f_{GLCM} + f_{HuM} + f_{Entropy}$

Thresholded image= $FV_2 = f_{DWT} + f_{GLCM} + f_{HuM} + f_{Entropy}$ (15)

ROI= $FV_3 = f_{DWT} + f_{GLCM} + f_{HuM} + f_{Entropy}$

### 3.4 Feature fusion

Features fusion is widely adopted in the Machine-Learning (ML) and Deep-Learning (DL) systems to enhance the classification accuracy. This practice is used to increase the size of the 1D FV to enhance the detection accuracy. In this work, the number of features existing in the considered FV is less (ie. 47x3=141). Hence, a serial fusion technique is employed to fuse the FVs, such as $FV_1$, $FV_2$ and $FV_3$.

The fused feature vectors considered in the proposed MLS is depicted below;

$$FFV_1 = FV_1 + FV_3 \tag{16}$$

$$FFV_2 = FV_1 + FV_2 + FV_3 \tag{17}$$

In which, the $FFV_1$ = 94x1 features, and $FFV_1$ =141x1 features.

### 3.5 Classifier implementation

In the ML and DL techniques, the classifiers are implemented to separate the given dataset into two or multi-class with the help of the feature-vector. Further, the choice of an appropriate classifier is essential to maintain the detection accuracy during the medial data assessment. In the proposed work, a two-class classification problem is considered and the implemented classifier is utilized to classify the CTS image dataset into normal/COVID-19 class.

In the proposed work, most commonly implemented classifiers, such as Naive Bayes (NB), k-Nearest Neighbors (KNN), Decision Tree (DT), Random Forest (RF) and Support Vector Machine (SVM) with linear kernel are employed to classify the considered images using the feature vectors, like $FV_1$, $FFV_1$ and $FFV_2$. The theoretical background of NB, KNN, DT, RF and SVM can be found in the literature [43-47, 50-53].

### 3.6 Performance validation

The eminence of ML and DL based data analysis is generally authenticated by calculating the important performance values. In the proposed MLS, the following performance values are computed to validate the eminence of the implemented classifier system.

The mathematical expression for performance values are as follows:

$$False\ Negative\ Rate = FN_r = \frac{F_{-ve}}{F_{-ve} + T_{+ve}} \tag{18}$$

$$False\ Positiveive\ Rate = FP_r = \frac{F_{+ve}}{F_{+ve} + T_{-ve}} \tag{19}$$

$$Accuracy = ACC = \frac{T_{+ve} + T_{-ve}}{T_{+ve} + T_{-ve} + F_{+ve} + F_{-ve}} \tag{20}$$

$$\Pr ecision = PRE = \frac{T_{+ve}}{T_{+ve} + F_{+ve}} \tag{21}$$

$$Sensitivity = SEN = \frac{T_{+ve}}{T_{+ve} + F_{-ve}} \tag{22}$$

$$Specicity = SPE = \frac{T_{-ve}}{T_{-ve} + F_{+ve}} \tag{23}$$

$$Negative\ Predictive\ Value = NPV = \frac{T_{-ve}}{T_{-ve} + F_{-ve}} \tag{24}$$

$$F1Score = F1S = \frac{2T_{+ve}}{2T_{+ve} + F_{-ve} + F_{+ve}} \tag{25}$$

where $F_{+ve}$, $F_{-ve}$, $T_{+ve}$, and $T_{-ve}$ represents, false-positive, false-negative, true-positive, and true-negative respectively.

### 3.7 COVID-19 dataset

The clinical level analysis of the pneumonia infection due to COVID-19 is generally assessed using CTS. This work considered 500 numbers of grayscale lung CTS (250 normal and 250 COVID-19 class) for the estimation. The normal CTS are collected from the LIDC-IDRI [54-56] and the RIDER-TCIA [57,58] and the COVID-19 class images are collected from the Radiopaedia database

[59-67] and the benchmark test images available at [68]. All these images are resized into 512x512x1 pixels and the resized images are considered for the experimental investigation. The sample test images considered in the proposed study is depicted in Figure 3.

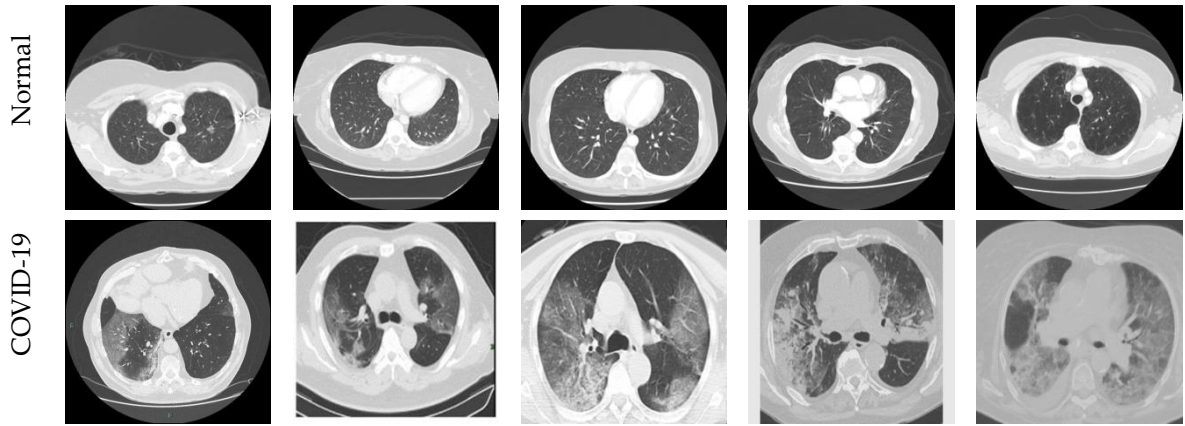

**Figure 3.** Sample test images considered in the proposed study

**4. Result and Discussion**

In this section, the investigational outcome achieved are presented and discussed. This MLS is implemented using a workstation with configuration-Intel i5 2.GHz processor with 8GB RAM and 2GB VRAM equipped with the MATLAB®. Experimental outcome of this MLS authenticate that it needs a mean time of 183±17sec to process the considered CTS dataset. The benefit of this MLS is, it is an automated technique and will not involve the operator assistance during the CTS classification.

The performance of implemented thresholding and segmentation technique is initially executed using the clinical grade CTS provided in Radiopaedia case-study [69] and the attained results are presented in Figure 3. In this work, the axial, coronal and sagittal slices of the case-study is assessed to confirm the performance of the proposed system and finally the infection due to COVID-19 is extracted using the watershed segmentation recently discussed in [26]. The results confirm that, proposed work offered better segmentation on the considered CTS irrespective of its orientation. Fig 4(a) and (b) depicts the test image to be evaluated and CBA+KE thresholded image respectively. Fig 4(c) and (d) depicts the outcome of the threshold filter, such as >*Th* and <*Th* respectively. Fig 4(e) presented the extracted infection using the watershed algorithm. From Fig 4(e), it can be noted that, proposed scheme works well on all the orientations of the CTS and extracts the infection with better accuracy.

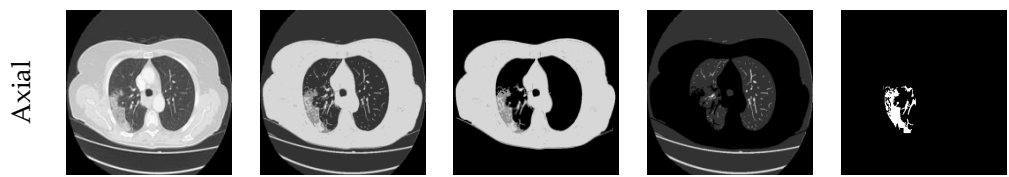

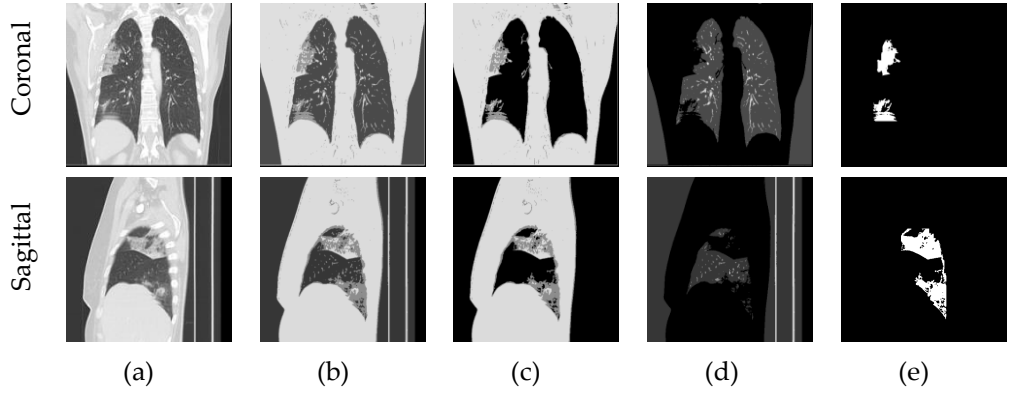

**Figure 4.** Sample results attained with 2D slices of the CT scan images.
(a) Test image, (b) Threshold image, (c) Separated section with threshold > Th,
(d) Extracted section with threshold < Th, (e) Extracted COVID-19 infection

After confirming the segmentation performance on the considered case-study, the proposed MLS is then considered to classify the CTS database into normal/COVID-19 class using a chosen procedure. As discussed in section 3, all the images of the considered CTS database are initially enhanced using the CBA+KE threshold and from the enhanced image the ROI is extracted by implementing the filter with a chosen thereshold of *Th=179±4*. Later, the essential image features, from the original, threshold image and the ROI are extracted using DWT, GLCM, HuM and entropies; as discussed in section 3.3. Later, the feature selection is implemented for the DWT, which helped to reach a final feature vector of dimension 47x1 is reached and is then named as $FV_1$, $FV_2$ and $FV_3$ The feature parameters in *FV₁* and *FV₂* are approximately similar (*FV₁≈FV₂*), hence, during the feature fusion task, the fused-feature-vector is attained as follows; $FFV_1 = FV_1 + FV_3$ and $FFV_2 = FV_1 + FV_2 + FV_3$. The classifier training, testing and validation is separately implemented using $FV_1$, $FFV_1$, and $FFV_2$; and the attained results are then analyzed to identify and confirm the best possible classifier for the proposed MLS.

Initially, $FV_1$ is considered to evaluate the performance of the classifier on the considered data and the attained results are depicted in Figure 5. The performance of the classifier is verified using a five-fold cross validation and the best value among the five trials are chosen for the assessment. Fig 5(a) and (b) shows the confusion-matrix obtained for the NB and KNN classifier. Fig 5(c) – (e) depicts the confusion-matrix of DT, RF and SVM respectively. From this confusion-matrix, it can be noted that the accuracy and overall performance offered by the SVL is superior compared to other classifiers, considered in this research. The classification accuracy attained by the SVM is 81.20%

|  | | Detected | | |
|---|---|---|---|---|
|  | | COVID-19 | Normal | |
| Actual | COVID-19 | $T_P$=203 | $F_N$=47 | SEN=0.8120 |
|  | Normal | $F_P$=52 | $T_N$=198 | SPE=0.7920 |
|  |  | PRE=0.7961 | NPV=0.8082 | ACC=0.8020 |

(a) NB

|  | | Detected | | |
|---|---|---|---|---|
|  | | COVID-19 | Normal | |
| Actual | COVID-19 | $T_P$=198 | $F_N$=52 | SEN=0.7920 |
|  | Normal | $F_P$=44 | $T_N$=206 | SPE=0.8240 |
|  |  | PRE=0.8182 | NPV=0.7984 | ACC=0.8080 |

(b) KNN

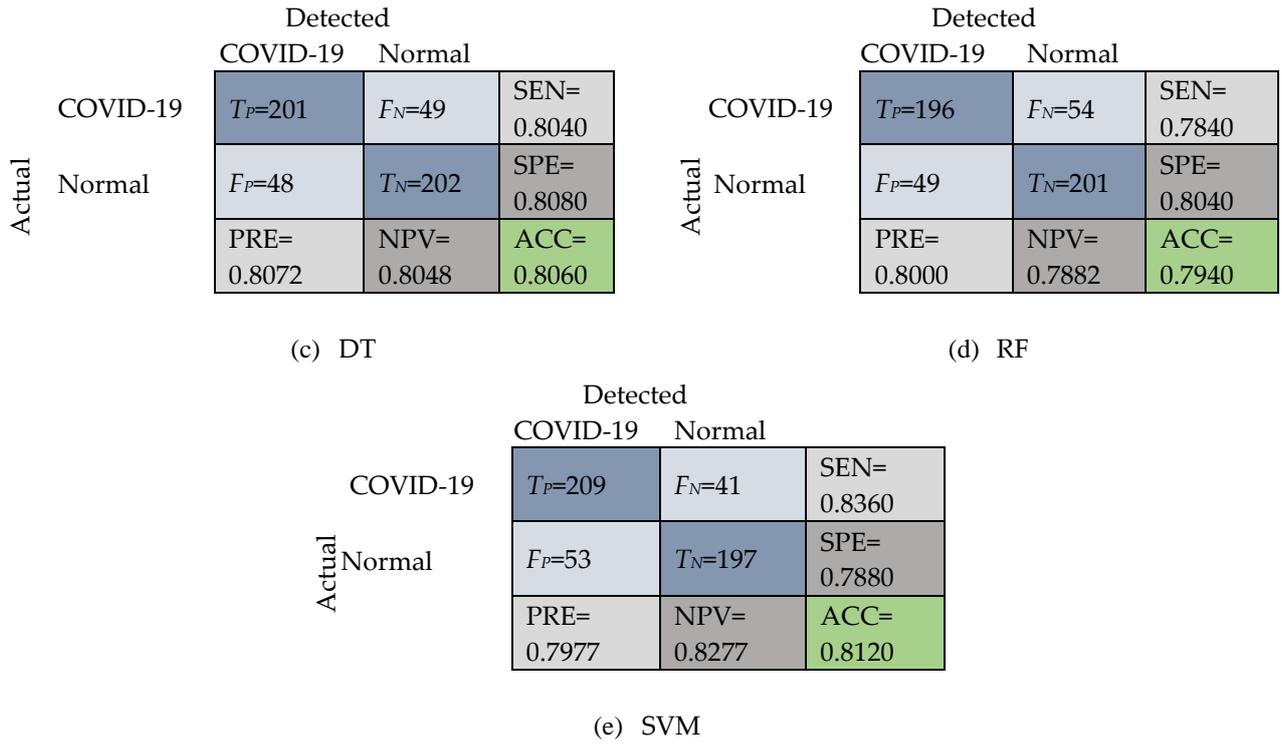

**Figure 5.** Confusion metrics achieved during the classification task executed with FV

Similar procedure is then repeated using $FFV_1$ as well as $FFV_2$ and the results attained are presented in Table 2 and Table 3. The results of these tables confirm that, the classifier accuracy achieved with the fused-feature-vector is better and these results confirm that, the increase in number of features will increase the classification accuracy. The classification accuracy attained with $FFV_2$ is better compared to the accuracy attained with $FFV_1$ as well as $FV_1$. Along with the accuracy, it is necessary to compute the overall accuracy of the classifier, to confirm its clinical significance. To get the information about the overall performance of the classifier, Glyph plot is considered in this work. The Glyph plot [] will provide a graphical representation based on the amplitudes of the performance measures. Usually, the Glyph plot with larger dimension represents the better overall performance and the Glyph plot achieved for various classifiers using $FV_1$, $FFV_1$, and $FFV_2$ are depicted in Figure 6.

From Fig 6(a) it can be noted that, the overall performance attained with SVM is superior. Fig 6(b) shows the better performance by RF and Fig 6(c) confirms the performance of the SVM. In the considered system, for all the feature cases, the overall performance attained with the classifiers, such as NB, KNN and DT are lesser compared to the RF and the SVM.

**Table 2.** Initial performance values achieved with the proposed system

| Features | | | Classifier | TP | FN | TN | FP | FNr | FPr |
|---|---|---|---|---|---|---|---|---|---|
| FVV1 | (FV1+FV3) | (94x1 features) | NB | 201 | 49 | 211 | 39 | 0.1960 | 0.1560 |
| | | | KNN | 207 | 43 | 209 | 41 | 0.1720 | 0.1640 |

|  |  | DT | 209 | 41 | 212 | 48 | 0.1640 | 0.1846 |
|  |  | RF | 216 | 34 | 213 | 37 | 0.1360 | 0.1480 |
|  |  | SVM | 212 | 38 | 211 | 39 | 0.1520 | 0.1560 |
| FVV2 (FV1+FV2+FV3) (141x1 features) |  | NB | 222 | 28 | 217 | 33 | 0.1120 | 0.1320 |
|  |  | KNN | 226 | 24 | 210 | 40 | 0.0960 | 0.1600 |
|  |  | DT | 224 | 26 | 218 | 32 | 0.1040 | 0.1280 |
|  |  | RF | 228 | 22 | 213 | 37 | 0.0880 | 0.1480 |
|  |  | SVM | 218 | 32 | 231 | 19 | 0.1280 | 0.0760 |

**Table 3.** Performance measures attained using the proposed machine-learning scheme

| Features | Classifier | ACC | PRE | SEN | SPE | F1S | NPV |
| --- | --- | --- | --- | --- | --- | --- | --- |
| FVV1 (FV1+FV3) (94x1 features) | NB | 0.8240 | 0.8375 | 0.8040 | 0.8440 | 0.8204 | 0.8115 |
|  | KNN | 0.8320 | 0.8347 | 0.8280 | 0.8360 | 0.8313 | 0.8294 |
|  | DT | 0.8255 | 0.8132 | 0.8360 | 0.8154 | 0.8245 | 0.8379 |
|  | RF | 0.8580 | 0.8538 | 0.8640 | 0.8520 | 0.8588 | 0.8623 |
|  | SVM | 0.8460 | 0.8446 | 0.8480 | 0.8440 | 0.8463 | 0.8474 |
| FVV2 (FV1+FV2+FV3) (141x1 features) | NB | 0.8780 | 0.8706 | 0.8880 | 0.8680 | 0.8792 | 0.8857 |
|  | KNN | 0.8720 | 0.8496 | 0.9040 | 0.8400 | 0.8760 | 0.8974 |
|  | DT | 0.8840 | 0.8750 | 0.8960 | 0.8720 | 0.8854 | 0.8934 |
|  | RF | 0.8820 | 0.8604 | 0.9120 | 0.8520 | 0.8854 | 0.9064 |
|  | SVM | 0.8980 | 0.9198 | 0.8720 | 0.9240 | 0.8953 | 0.8783 |

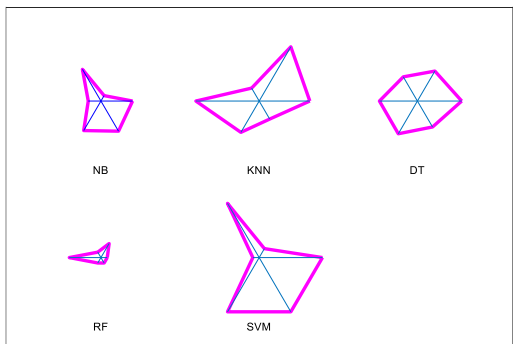

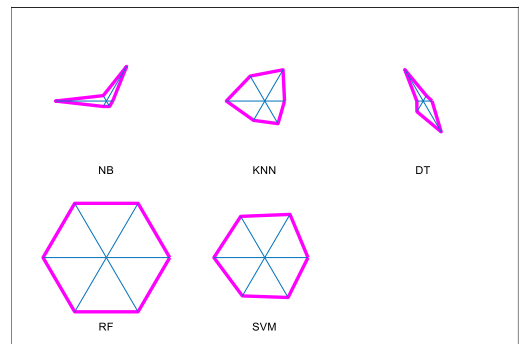

(a)            (b)

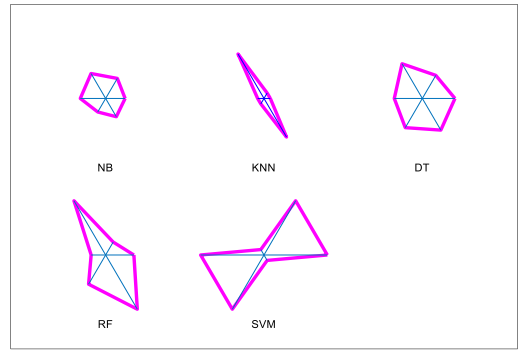

(c)

**Figure 6.** Glyph plot of the performance measures obtained during the classification task.

(a) Classification with $FV_1$, (b) Classification with $FFV_1$ (c) Classification with $FFV_2$

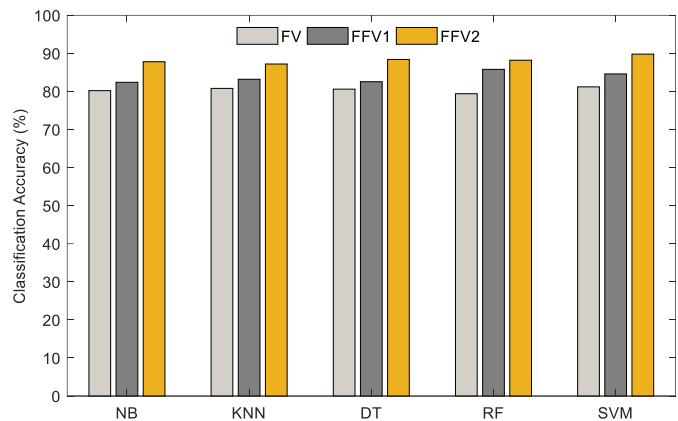

**Figure 7.** Classification accuracy achieved with the proposed system with various feature vectors

The results presented in Figure 7 confirm that, when the number of features is increased, then the classification accuracy can be increased. Further, the accuracy attained using $FFV_2$ is superior compared with the accuracy attained with other irrespective of the classifier unit.

In the proposed research, the MLS is implemented to examine the CTS dataset of normal/COVID-19 class and attained an accuracy of >89% with the SVM classifier. In future, a suitable DL procedure can be implemented to improve the classification accuracy.

**5. Conclusions**

The aim of this research is to propose a computerized system to distinguish the normal and COVID-19 CTS images from a considered image database. This work proposes a MLS using a sequence of procedures ranging from image pre-processing to the classification to implement a scheme with better detection accuracy. The proposed MLS initially implements an image thresholding process with CBA+KE to enhance the test image and then implements a threshold filter to separate the ROI and artifact. Later, essential procedures, such as feature-extraction, feature-selection, feature-fusion, and classification are employed in the proposed MLS. In this work the classifier units, like NB, KNN, DT, RF and SVM are considered and its performance are individually tested with chosen features, such as $FV_1$, $FFV_1$ and $FFV_2$. The experimental investigation of this study confirms that, the classification accuracy of SVM is 89.80% when $FFV_2$ is considered to train, test and validate the classifier. This confirms that, when the proposed MLS is equipped with the SVM classifier, a better classification is attained with the considered CTS database.